\documentclass[runningheads,dvipsnames]{llncs}
\usepackage{amsmath}

\usepackage{graphicx}
\usepackage{verbatim}

\usepackage{tikz}
\usetikzlibrary{calc}
\usetikzlibrary{arrows.meta}
\usepackage{graphicx,adjustbox}
\usepackage{pgfplots}
\usepgfplotslibrary{groupplots}

\usepackage{hyperref}
\usepackage{caption}
\usepackage{subcaption}

\usepackage{multicol}

\usepackage{multirow}
\usepackage{algorithm,algpseudocode}
\usepackage{listings}
\usepackage{amsmath}
\newcommand{\pushcode}[1][1]{\hskip\dimexpr#1\algorithmicindent\relax}
\MakeRobust{\Call}
\makeatletter
\renewcommand*{\ALG@name}{Listing}
\makeatother
\begin{document}
\title{Implementing Window Functions in a\\ Column-Store with Late Materialization (extended version)}
\titlerunning{Window Functions in Column-Store}
%
\author{Nadezhda Mukhaleva\inst{2}\orcidID{0000-0002-8552-9274} \and
Valentin Grigorev\inst{1,2}\orcidID{0000-0003-4235-3712} \and
George Chernishev\inst{1,2}\orcidID{0000-0002-4265-9642}\\
\{nmukhaleva,valentin.d.grigorev,chernishev\}@gmail.com}
\authorrunning{N. Mukhaleva et al.}
%
\institute{Information Systems Engineering Lab, JetBrains Research\\ \url{https://research.jetbrains.org/groups/information\_lab}  \and   Saint Petersburg University, Russia\\ \url{http://english.spbu.ru/}}
\maketitle              
\begin{abstract}

A window function is a generalization of the aggregation operation. Unlike aggregation, the cardinality of its output is always the same as the cardinality of input. That is, the semantics of this operator imply computing values for extra attributes for each row, depending on its context, either expressed by a sliding window or a previously evaluated row. Window functions are a very powerful tool, which is also popular among data analysts and supported by the majority of industrial DBMSes. It allows to gracefully express quite complex use-cases, such as running sums and averages, local maximum and minimum, and different types of ranking. Since they can be expressed without self-joins and correlated subqueries, their evaluation can be performed much more efficiently.

In this paper we discuss an implementation of window functions inside a disk-based column-store with late materialization. Late materialization is a technique that aims to keep tuple reconstruction back from individual columns as long as possible. Initially popular in the late 00's, it is rarely considered nowadays. However, in case of window functions it allows to substantially lower memory footprint. Another contribution of this paper is the application of a segment tree to computing RANGE-based window functions. 

\keywords{Window Function \and Analytical Function \and Aggregation \and Column-Store \and Query Processing \and Late Materialization \and OLAP \and PosDB}
\end{abstract}

\section{Introduction}\label{sec:introduction}

\footnotetext{This is an extended version of the paper published in the proceedings of MEDI'19.}

A column-store is a type of DBMS designed specifically for handling analytic applications. Its core idea is to store each attribute separately, either on disk or in memory. This type of storage allows to implement the so-called lightweight compression schemes~\cite{Abadi:2006:ICE:1142473.1142548,columns_tutorial} efficiently, due to the resulting data homogeneity. Column-stores both store and operate on data in columnar form. The utilized approach to processing can be used to classify column-store systems~\cite{Abadi:2013:DIM:2602024}: 

\begin{itemize}
    \item ``Naive'' column-stores. In these systems, data is stored and processed in columnar form only on the lowest levels of the operator tree~\cite{graefe_query_1993}. Usually, it happens as follows: each column is read, decompressed and filtered. Then, all columns corresponding to attributes of a single table are ``glued'' together (i.e. tuple reconstruction is preformed), after which processing continues similarly to a row-store.
    \item Full-fledged column-stores. These systems feature a so-called late materialization approach. In this case, tuple reconstruction is delayed to the latest possible time, and until this moment the system operates on positions.
\end{itemize}

The majority of existing column-stores are ``naive'', there is only a handful of systems that support late materialization~\cite{Abadi:2013:DIM:2602024}. Even the ``naive'' approach has allowed column-stores to beat classic systems in terms of query processing speeds. In its turn, late materialization allows to achieve even better performance. 

In the 00's, late materialization received a great deal of attention since it was one of the strong points of column-stores. Nowadays, this interest had largely faded away, mainly due to the complexity of implementation~\cite{6544909}. Nevertheless, late materialization looks  promising for evaluating aggregation~\cite{TuchinaEtAl:SEIM2018}, as well as for queries containing window functions.

Window function (or analytic function) is a concept that was proposed in reference~\cite{oraclereport} and later became a part of the SQL:2003 standard. This operation possesses the following semantics: similarly to aggregation, data is partitioned into several groups. Next, a sort may be applied to data from each group. The third step depends on the specified window functions. One of the classes uses framing, which is a process described as follows: 1) a sliding window is determined for each row, 2) an aggregation is performed for the data in the window,  3) the result is ``added'' to the current row. Another class operates on an entire group: for each row it computes new values using the ones of the previously processed row (e.g. \texttt{RANK}).

Window functions are a very powerful tool, which is also popular among data analysts and supported by the majority of industrial DBMSes. 


In this paper, we discuss the implementation of window functions inside a column-store with late materialization that supports on-demand reading of individual columns. Employing late materialization may allow to speed up processing of such queries and reduce memory footprint, which is very important for this operation. We validate our approach by implementing it inside PosDB~\cite{chernishev_posdb:_2018,10.1007/978-3-319-74313-4-7}~--- a distributed disk-based column-store with late materialization~--- and by comparing it with PostgreSQL. 

Overall, the contribution of this paper is the following:
\begin{enumerate}
    \item An adaptation of the window function processing approach for column-stores with late materialization. We present three different strategies which are sufficiently generalized to be implemented in any column-store that allows per-attribute data reading.
    \item A model for estimating memory requirements for each of these strategies.
    \item An enhancement of the segment tree technique for processing \texttt{RANGE}-based window functions. 
\end{enumerate}

This paper is organized as follows: in Section~\ref{sec:background} we provide the basic knowledge necessary for further understanding of this paper. In particular, the semantics and syntax of window functions are described and PosDB architecture is presented. Next, in Section~\ref{sec:wf_processing} we describe existing approaches to evaluating window functions. Section~\ref{sec:proposed_approach} contains the contribution of this paper. In Section~\ref{subsec:late_classic} a description of possible strategies to implementing window functions in late materialization environment and a model for estimating memory requirements are provided. Section~\ref{subsec:st_range} describes an enhancement of a segment tree technique for \texttt{RANGE}-based window functions. Sections~\ref{sec:experiments}, \ref{sec:related_work}, \ref{sec:conclusion} present experiments, related work, and conclusion, respectively.

\section{Background}\label{sec:background}

\subsection{Window Functions Basics}\label{subsec:wf_basics}


The PostgreSQL documentation~\cite{PostgreSQLDoc} states that ``a window function performs a calculation across a set of table rows that are somehow related to the current row''. This set of rows represents the context of calculation. Traditionally, it is called a window or a frame. Evaluation of window functions is performed after completion of most operations, but before sorting and duplicate elimination.


Window function calls are placed in the \texttt{SELECT} clause of the query and have the following structure:

\begin{lstlisting}
window_function(column) OVER (
    [ PARTITION BY column [, ...] ]
    [ ORDER BY { column [ ASC | DESC ] } [, ...]  ]
    [ { ROWS | RANGE } BETWEEN frame_start AND frame_end ]
),
\end{lstlisting}
\noindent where \texttt{frame\_start} may be:
\begin{itemize}
    \item \texttt{UNBOUNDED PRECEDING}
    \item \textbf{\textit{offset}} \texttt{PRECEDING}
    \item \texttt{CURRENT ROW}
\end{itemize}
\noindent and \texttt{frame\_end} may be: 
\begin{itemize}
    \item \texttt{CURRENT ROW}
    \item \textbf{\textit{offset}} \texttt{FOLLOWING}
    \item \texttt{UNBOUNDED FOLLOWING}
\end{itemize}

Generally, \textbf{\textit{offset}} can be a function of the current row values, but this case is quite uncommon. It is not implemented in the majority of DBMSes. 

If several window functions appear in the same context, then window definition should be moved to the end of the query and placed into the \texttt{WINDOW \textit{window\_name} AS (\textit{window\_definition})} statement. In this case, attributes appearing in the \texttt{SELECT} clause that require window functions should reference the respective definitions by name.

Processing of window functions can be split into steps that are closely related to the corresponding syntax constructions. During the first step~--- partitioning~--- input data is divided into several groups based on the attribute values mentioned in the \texttt{PARTITION BY} clause. It is done similarly to classic aggregation, but in aggregation we often can fold\footnote{Traditional higher order function \texttt{reduce} or \texttt{fold} is meant here.} data on the fly, but in this case rows should be stored in their original form. Regarding the implementation, it can be done using either a hash table or sorting. In general case, sorting is a more time-consuming operation, so in this paper we assume that partitioning is implemented with hashing.

The next step is the ordering, which is described by the \texttt{ORDER BY} clause of window definition. For each group, it imposes a certain order over tuples. In general, this step is optional, but for some operations, i.e. ranking functions and \texttt{RANGE}-based window functions, it is mandatory.


The last step of window function processing is its evaluation.  Despite the name, not all window functions work with the window, e.g., for ranking functions, the window is defined as the whole group. Nevertheless, such window functions assume presence of a partitioning and an in-group ordering similarly to ``true'' window functions. Both types of window functions have the same semantics: cardinality of the output is the same as the cardinality of the input. In other words, for each considered row processing only ``appends'' additional attributes. For this reason both types are usually considered together when window function evaluation is discussed.

``True'' window functions run aggregation over the frame formed around the current row according to a rule specified by the \texttt{OVER} clause. If the \texttt{ROWS} framing is used, then the offset denotes the number of rows to include before and after the current row.  If the \texttt{RANGE} framing is used, then it is strictly required that groups should be sorted by the attribute for which we evaluate the window function\footnote{i.e., this attribute should be first in the \texttt{ORDER BY} clause.}. In this case, the frame contains values from the range $[\mathit{current\_value} - \mathit{offset}_1; \mathit{current\_value} + \mathit{offset}_2]$. In both cases the start of the window can be tied to the first row of the group, and the end of the window can be tied to the last row of the group.

\subsection{PosDB Basics}\label{subsec:posdb_basics}

Our study of window functions is conducted inside PosDB~--- a distributed disk-based column-store. A detailed discussion of its architecture can be found in papers \cite{chernishev_posdb:_2018,10.1007/978-3-319-74313-4-7}. Here, we only describe the most important points. 

Query processing in PosDB is organized according to the pull-based Volcano model~\cite{graefe_query_1993} with block-oriented processing. This model assumes that query plans are represented by trees with operators as nodes and data flows as edges. Each operator supports an iterator interface and can produce either positions or tuples, exclusively. Currently, PosDB supports only late materialization, but a new model that will enable to express different materialization strategies is being actively developed. 

\begin{figure}
\centering
\begin{subfigure}{.3\textwidth}
    \centering
    \includegraphics[width=0.9\textwidth]{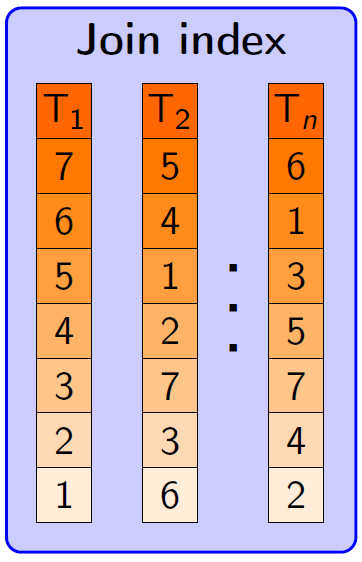}
    \caption{Example of join index}
    \label{fig:join-index}
\end{subfigure}%
\begin{subfigure}{.7\textwidth}
    \centering
    \includegraphics[width=0.9\textwidth]{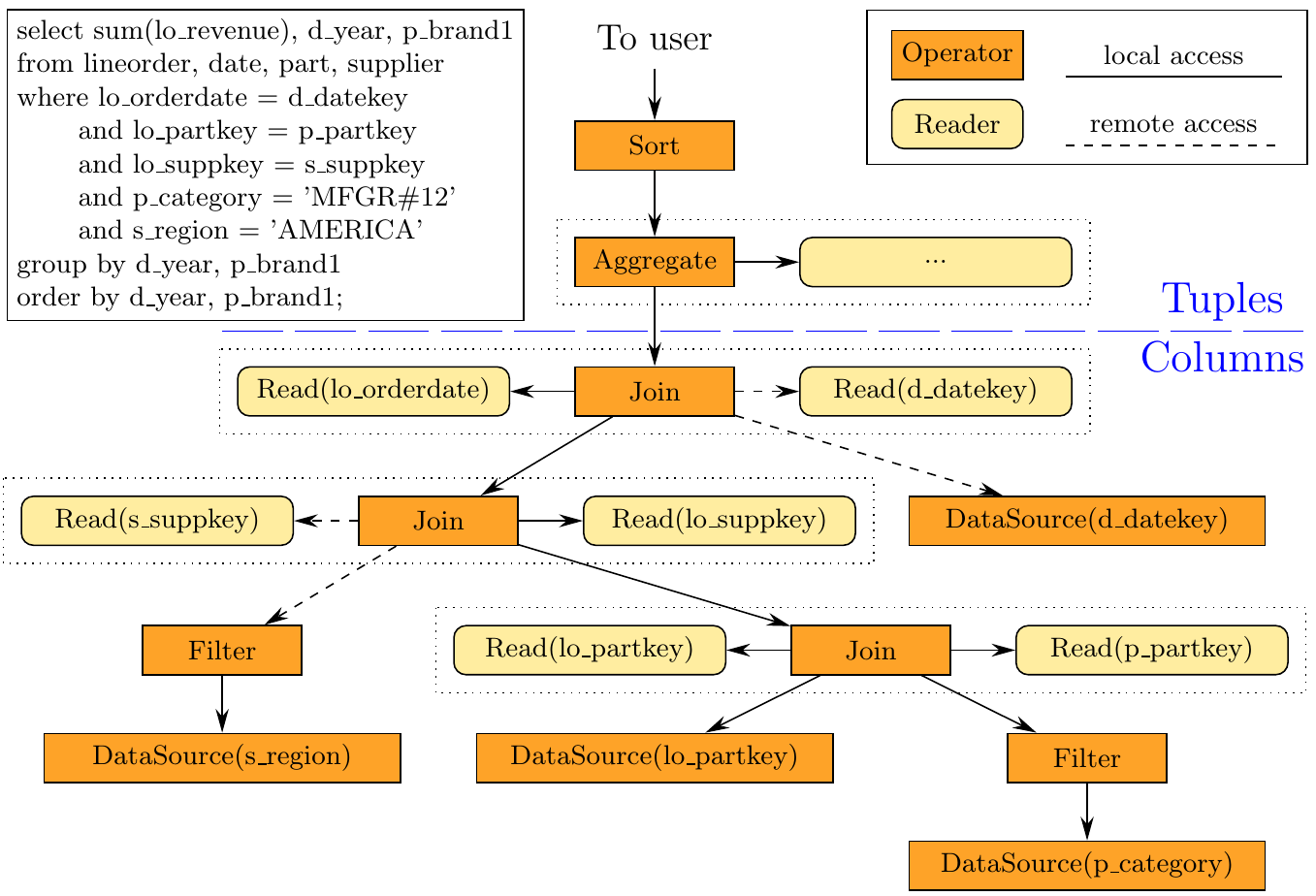}
    \caption{Query plan example}
    \label{fig:plan}
\end{subfigure}
\caption{PosDB internals}
\label{fig:test}
\end{figure}

To represent intermediate positional data PosDB uses a generalized join index~\cite{Valduriez:1987:JI:22952.22955}, a data structure that essentially encodes the result of a series of joins. The join index states that row $r_1$ of table $T_1$ was joined with row $r_2$ of table $T_2$ and so on. An example is presented in Figure~\ref{fig:join-index}. Past the materialization point, data is represented by tuples, similarly to row-stores.

Currently, in PosDB query plans consist of two sections: positional (columnar) and tuple parts. Operators belonging to the first one (joins, filter, positional bitwise operators) use a join index to represent intermediate results. The tuple part (aggregation and sort operators) is similar to row-stores. An example plan is presented in Figure~\ref{fig:plan}, where the dashed line denotes the materialization point.

Several operators like the positional bitwise \texttt{AND} and \texttt{OR} only need positional data, while others, i.e. join, filter, and aggregation also require the corresponding values. To fetch values using positions from join index, we have introduced special entities named readers. In this work, it is sufficient to distinguish two types of readers~--- \texttt{ColumnReader} and \texttt{SyncReader}. \texttt{ColumnReader} is designed to retrieve values of a specific attribute, and \texttt{SyncReader} is designed to read values of several attributes synchronously.

\section{Window Function Processing: Approaches and Algorithms}\label{sec:wf_processing}

In this section we discuss how to design the operator for window function evaluation. A comprehensive overview of existing approaches can be found in the article~\cite{leis_window_functions_2015}.
Overall, \texttt{Window Operator} can be implemented using two algorithms: the classic and the one based on a segment tree. These algorithms have a significant common part which is as follows: at first, partitioning and in-group ordering is performed. Next, groups are iterated over and each of them is processed independently of others. The distinction between these two algorithms is the group processing itself. The classic algorithm goes through tuples belonging to a single group while performing the following. For each tuple it computes frame bounds and then evaluates the window function over the data belonging to the frame. There are two possible approaches to this: naive and cumulative. The naive approach is straightforward: it calculates data from scratch for every frame instance. On the other hand, the idea of cumulative approach is to store the results of processing of the previous frame and to reuse them to evaluate current frame faster. It is very efficient in case of the \texttt{SUM} window function: the result for the previous frame is saved and using only one or two (one if any border of the frame is fixed and two otherwise) arithmetic operations allows to obtain a result for the current frame\footnote{Note that it is assumed here that frame offset does not depend on the current row value. Otherwise, the cumulative approach is still attractive, but not as dramatically.}. 

In case of the \texttt{MIN} and \texttt{MAX} window functions, the cumulative approach can be implemented by preserving the previous frame using a binary search tree. This is not as promising, but still can be useful, especially for large windows. It is straightforward to find window bounds defined by the \texttt{ROWS} clause while for the \texttt{RANGE} one they can be found using binary search.

As it was already mentioned earlier, another way of window processing is based on the segment tree data structure~\cite{emaxx:segment_tree}. This approach is relatively novel: it has been proposed in 2015 in the paper~\cite{leis_window_functions_2015}. This approach is as follows: at first, a segment tree is created from group data, and then tuples are iterated over. However, instead of computing over the current frame, a request is issued to the segment tree. This approach allows to efficiently evaluate window functions that have frame borders depending on the current row and to implement intra-group parallelism. 

In the original paper, it was considered only for the \texttt{ROWS} framing. In our paper, we propose a slightly generalized segment tree that can be utilized for \texttt{RANGE}-based window functions too. Details of this generalization are described in the Section~\ref{subsec:st_range}.
 
Window functions that do not require framing can be evaluated with a simplified version of the classic algorithm. 
 
\section{Proposed Approach}\label{sec:proposed_approach}

\subsection{Adapting Classic Algorithm for PosDB}\label{subsec:late_classic}

While in row-stores and column-stores with early materialization, the algorithm of the \texttt{Window Operator} is defined quite clearly, systems with late materialization can offer a variety of options. The variations are largely associated with the point of materialization inside the operator.

Recall that in PosDB, every query plan consists of two parts: positional- and tuple-oriented. If \texttt{Window Operator} is located after aggregation in a query plan, then its inputs are tuple blocks, and evaluation can be performed by one of previously described algorithms without any changes.

But if \texttt{Window Operator} receives positional data, we have to decide when tuples should be materialized. This task is not as simple as it may seem. The following variants are possible:
\begin{enumerate}
    \item \textbf{Strategy~1}. Tuples are materialized during hash table population. All of the subsequent stages of the algorithm are identical to the row-store case.
    \item Only keys are materialized during hash table population, and positional data is stored as values. In some cases, this allows to significantly reduce the size of the hash table. It is important to emphasize that the ordering step (see Section~\ref{subsec:wf_basics}) ceases to be a separate step of processing and becomes a part of the evaluation step. Thus, for each group, processing should start with the ordering. Further steps can be done in a number of ways:
    \begin{enumerate}
        \item \textbf{Strategy~2a}. At the beginning of group processing, all required attributes are materialized. Afterwards, tuples are sorted and window functions evaluation is performed as usual. Tuples are materialized only for one group at a time.
        \item \textbf{Strategy~2b}. At the beginning of group processing, only attributes required for ordering are materialized and ordering is performed. After this we can move through associated positions and materialize data on demand. This strategy is not implemented yet since it requires a new execution model for efficient implementation, but still, it looks quite promising for window functions over a fixed-size frame.
    \end{enumerate}
\end{enumerate}

It is reasonable to use \textbf{Strategy~1} if positions received by the \texttt{Window Operator} are ordered, since corresponding values can be read by a sequential scan. For example, such situation occurs when window functions are evaluated on a single table, i.e. the query does not contain joins. In other cases all these strategies require an equal number of I/O operations, so there should be no significant difference between them in processing time. At the same time, the amount of required memory can vary substantially.

Let us estimate the amount of memory required by all these strategies. It is necessary to introduce several variables and functions for estimation:
\begin{itemize}
    \item $A$~--- a set of all attributes which have to be materialized in some way;
    \item $A_k$~--- a set of partitioning attributes;
    \item $A_\mathit{sort}$~--- a set of sorting attributes;
    \item $A_\mathit{aggr}$~--- a set of attributes for which window functions are being evaluated\footnote{In our implementation, several window functions can be processed at once if they are defined over the same window};
    \item $N$~--- number of logical rows in the input;
    \item $G_{\mathit{key}}$~--- group corresponding to partitioning key  $\mathit{key}$ as a list of logical rows;
    \item $|G_\mathit{key}|$~--- number of logical rows in the group $G_\mathit{key}$;
    \item $G^\mathit{max} = \underset{G_\mathit{key}}{\mathrm{arg\,max}}\ {|G_\mathit{key}|}$
    \item $M$~--- number of groups;
    \item function $\operatorname{size}_t$~--- returns size of tuple from the corresponding set of attributes;
    \item function $\operatorname{size}_p$~--- returns size of logical row of positions for corresponding set of attributes; actually it is determined by the amount of tables joined before \texttt{Window Operator}.
\end{itemize}

In \textbf{Strategy~1}, materialized data is stored in the hash table and processing is run directly on it. It requires $$ \overbrace{M \times \operatorname{size}_t(A_k)}^\text{hash table keys} + \overbrace{N \times \operatorname{size}_t (A_\mathit{sort} \cup A_\mathit{aggr})}^\text{hash table data}.$$

In \textbf{Strategy~2a}, only tuples for keys are materialized during hash table population. The data itself is stored in the positional representation. Other attributes are being materialized during group processing, while dynamically deleting already read positions. Thus, this strategy requires
\begin{align*}
    \overbrace{M \times \operatorname{size}_t(A_k)}^\text{hast table keys} + &\overbrace{N \times \operatorname{size}_p(A_\mathit{sort} \cup A_\mathit{aggr})}^\text{hash table data} +\\ &\overbrace{G^\mathit{max} \times \Big(\operatorname{size}_t(A_\mathit{sort} \cup A_\mathit{aggr}) - \operatorname{size}_p(A_\mathit{sort} \cup A_\mathit{aggr})\Big)}^\text{$\Delta$ for  materialization with dropping processed positions}.
\end{align*}

Utilizing this strategy to process several window functions with the same window but over different attributes can result in significant performance improvement. 

\textbf{Strategy~2b} is a further enhancement of the same idea. Here, on the group processing stage only sorting attributes are materialized and thus, even better results are obtained:
\begin{align*}
    \overbrace{M \times \operatorname{size}_t(A_k)}^\text{hash table keys} + &\overbrace{N \times \operatorname{size}_p(A_\mathit{sort} \cup A_\mathit{aggr})}^\text{hash table data} +\\ &\overbrace{G^\mathit{max} \times \operatorname{size}_t(A_\mathit{sort})}^\text{materialized sorting attributes} + \overbrace{\operatorname{size}_t(A_\mathit{aggr}) \times \mathit{window\_size}}^\text{other attributes materialized for the window}
\end{align*}

\subsection{RANGE-based window functions}\label{subsec:st_range}

At first, let us discuss implementation details of the segment tree data structure~\cite{emaxx:segment_tree}. In literature, it is usually described for specific operations, such as \texttt{SUM}, \texttt{MIN} or \texttt{MAX}. But here, a general solution is required and this leads to remarkable implications.  

It is common to implement the segment tree on the base of an array with an implicit tree structure, since the segment tree is always a complete tree. 
For a complete tree, an array is the most space-efficient representation\footnote{In an array-based implementation, we store just data without auxiliary information such as pointers to children which are necessary to describe an arbitrary binary tree.}. 

However, an issue arises: if the tree is not a perfect\footnote{\url{https://xlinux.nist.gov/dads//HTML/perfectBinaryTree.html}} binary tree (i.e. the last level is not completely filled, or, in other words, the original array size is not a power of $2$), then some cells in the array are left uninitialized. Usually, when a segment tree data structure is being discussed, it is considered in a form tuned for a particular operation. Here, however, it is necessary to consider a number of operations. First of all, in order to properly initialize this array, an identity element should be chosen. Obviously, it depends on the operation: $0$ for \texttt{SUM}, $-\infty$ for \texttt{MAX}, $+\infty$ for \texttt{MIN}, etc. 

It is reasonable to not store a chunk of the tree that consists only of identity elements. Instead, it is convenient to``overload'' the access to tree elements and return a ``virtual'' value if an out-of-real-bounds element is requested. 

Furthermore, it is easy to see that employing a segment tree leads to reordering of the sequence of operations. Thus, it is necessary to require associativity of the operation for which the segment tree is built. As the result, a segment tree requires an underlying data type with the corresponding operation to be a monoid~\cite{Jacobson1951}.

It is quite obvious that having a tuple of monoids, we can create a new monoid which works with tuples:

\begin{equation*}
  \begin{split}
    m_1 &= \{T_1, op_1:\ T_1 \times T_1 \rightarrow T_1 \}\\
    \ldots\\
    m_n &= \{T_n, op_1:\ T_n \times T_n \rightarrow T_n \}
  \end{split}
\quad\Longrightarrow\quad
  \begin{split}
    M &= \{ T = (T_1, \ldots, T_n), op:\ T \times T \rightarrow T \}
  \end{split}
\end{equation*}

This approach allows to efficiently process queries with several different window functions defined over the same \texttt{OVER} clause in case of \texttt{ROWS} framing.

For the \texttt{RANGE} framing, it is only reasonable to utilize this approach if several window functions over the same attribute need to be evaluated.  

Next, having discussed data organization in the segment tree, we are going to consider data processing. The construction of a segment tree is quite straightforward. A detailed description of this process can be found in the reference~\cite{emaxx:segment_tree}. 

Now, let us consider processing of queries in a segment tree. It is performed via a recursive function \texttt{evaluateSegment}, whose pseudocode and description are given later. To correctly start it, the  \texttt{evaluateFrame} function shown in the Listing~\ref{algo:evaluate_frame} is used. It initiates a recursive function on the root of tree that has the current segment covering the whole bottom level of the tree.

\begin{algorithm}
    \caption{Recursion initialization}
    \label{algo:evaluate_frame}
    \begin{algorithmic}
        \Function{evaluateFrame}{monoid, fLeft, fRight}
            \State \Return \Call{evaluateSegment}{0, monoid, 0, $2^{\lceil \log_2{ \mathit{array\_size}}\rceil}$ - 1, fLeaft, fRight}
        \EndFunction
    \end{algorithmic}
\end{algorithm}

Next, consider the classic recursive algorithm shown in Listing~\ref{algo:classic_segment_tree}. As input, it receives the current position in the tree \texttt{index}, a monoid (data type, identity, and operation), segment borders corresponding to the current node \texttt{cLeft} and \texttt{cRight}, and borders of the requested frame \texttt{fLeft} and \texttt{fRight}. Its output is a set of requested values over the specified frame. The algorithm itself is very simple: if current segment is correct and is not equal to the requested frame, then we split it in half and transfer control to children recursively. In this listing, monoid.op is the corresponding operation of monoid, e.g. \texttt{SUM}, \texttt{MAX}, etc.

\begin{algorithm}
    \caption{Classic \texttt{evaluateSegment} algorithm}
    \label{algo:classic_segment_tree}
    \begin{algorithmic}
        \Function{evaluateSegment}{index, monoid, cLeft, cRight, fLeft, fRight}
            \If{fLeft $>$ fRight}
                \State \Return monoid.identity
            \EndIf
            \If{fLeft = cLeft \textbf{and} fRight = cRight}
                \State \Return \Call{getValue}{index}
            \EndIf
            \State $\text{m} \gets (\text{cLeft} + \text{cRight}) / 2$
            \State \Return monoid.op(\\
                \pushcode[0] \hspace{1em} \Call{evaluateSegment}{$2 \cdot \text{index} + 1$, cLeft, m, fLeft, \Call{min}{fRight, m}},\\ 
                \pushcode[0] \hspace{1em} \Call{evaluateSegment}{$2 \cdot \text{index} + 2$, $\text{m} + 1$, cRight, \Call{max}{fLeft, $\text{m} + 1$}, fRight}\\
            \pushcode[0])
        \EndFunction
    \end{algorithmic}
\end{algorithm}
To support \texttt{RANGE}-based window function processing with segment tree, we have to generalize the algorithm described above. The interface modification is straightforward~--- we have to replace integer-valued \texttt{fLeft} and \texttt{fRight} with parameters corresponding to the processed attribute type. Also, several additional functions and variables have to be defined:
\begin{itemize}
    \item \texttt{nLeaves}~--- number of existing leaves on the bottom level, equal to the size of original array;
    \item \texttt{getLeafValue} function~--- get $k$-th element of the bottom level;
    \item \texttt{getLeafOrMax} function~--- call \texttt{getLeafValue} if index corresponds to existing value and return the last element of the bottom level if index is out-of-range.
\end{itemize}

The idea of recurrent tree traversal remains largely the same, but several changes are introduced. Firstly, we return the identity element if the left bound of the frame comes out of bounds. In comparison to the classic algorithm, it is necessary to explicitly check this. Furthermore, an equality check between the current segment and the requested frame should be replaced with an inclusion check, since it is a part of \texttt{RANGE} behavior. Furthermore, all comparisons require to wrap the current segment borders in \texttt{getLeafOrMax} calls. The resulting algorithm is presented in Listing~\ref{algo:range_segment_tree}.

\begin{algorithm}
    \caption{\texttt{evaluateSegment} algorithm for value ranges}
    \label{algo:range_segment_tree}
    \begin{algorithmic}
        \Function{evaluateSegment}{index, monoid, cLeft, cRight, fLeft, fRight}
            \If{fLeft $>$ fRight \textbf{or} cLeft $>=$ nLeaves}
                \State \Return monoid.identity
            \EndIf
            
            \If{fLeft $<=$ \Call{getLeafValue}{cLeft} \textbf{and}\\
                \pushcode[0]\hspace{2em} fRight $>=$ \Call{getLeafOrMax}{cRight}}
                \State \Return \Call{getValue}{index}
            \EndIf
            
            \State $\text{m} \gets (\text{cLeft} + \text{cRight}) / 2$
            \State \Return monoid.op(\\
                \pushcode[0] \hspace{1em} \Call{evaluateSegment}{$2 \cdot \text{index} + 1$, cLeft, m, fLeft, \Call{min}{fRight, \Call{getLeafOrMax}{m}}},\\ 
                \pushcode[0] \hspace{1em} \Call{evaluateSegment}{$2 \cdot \text{index} + 2$, $\text{m} + 1$, cRight, \Call{max}{fLeft, \Call{getLeafOrMax}{$\text{m} + 1$}}, fRight}\\
            \pushcode[0])
        \EndFunction
    \end{algorithmic}
\end{algorithm}

The introduced changes are very straightforward. Nevertheless, the authors of paper~\cite{leis_window_functions_2015} where the segment tree based algorithm was suggested did not consider the \texttt{RANGE} case. It is rather peculiar since the \texttt{RANGE} case looks inherently more suitable for processing with the segment tree based algorithm. \texttt{ROWS} framing with ``floating'' borders is very rare, while \texttt{RANGE} features such borders by its definition.

\section{Experiments}\label{sec:experiments}

Experimental evaluation was performed on a PC with the following characteristics: 4-core Intel\textregistered Core\texttrademark\ i5-7300HQ CPU @ 2.50GHz, 8 GB RAM, running Ubuntu Linux 18.04.2 LTS. We have used PostgreSQL 11.3 as a baseline for comparison. For our experiment, we have constructed the query template shown below. It is based on the \texttt{LINEORDER} table from the SSB benchmark~\cite{SSB}.

\begin{lstlisting}[language=sql,label={listing:query},basicstyle=\small]
SELECT lo_orderpriority, SUM(lo_ordtotalprice) OVER (
  PARTITION BY lo_orderpriority ORDER BY lo_ordtotalprice 
  RANGE BETWEEN @offt PRECEDING AND @offt FOLLOWING) AS sum 
FROM lineorder ORDER BY lo_orderpriority ASC
\end{lstlisting}

\noindent where \texttt{@offt} varies in range of [10, \ldots , 10M].

This experiment is quite simple and only demonstrates the attractiveness of a segment tree-based approach for processing of \texttt{RANGE}-based window functions. We believe that for a detailed performance analysis, a special benchmark has to be developed.

The aforementioned queries were run on PosDB and PostgreSQL with the SSB scale factors 1--7. The results of our experiments are presented in the table.

\begin{figure}[!htb]
    \centering
    \begin{minipage}{.5\textwidth}
        \centering
	\begin{tabular}{|c|c|c|c|c|c|}
		\hline
		\texttt{@offt} & DBMS & SF=1 & SF=3 & SF=5 & SF=7\\
		\hline\hline
		\multirow{2}{*}{10}& PosDB & $10611$ & $32888$ & $57484$ & $84935$\\
						   & Postgres  & $11498$  & $37205$ & $73003$ & $116160$\\\hline\hline 	
		\multirow{2}{*}{100}& PosDB & $11454$ & $35979$ & $63658$ & $93743$ \\
 				            & Postgres  & $11536$  & $38046$ & $65834$ & $111100$\\\hline\hline
 		\multirow{2}{*}{1K}& PosDB & $11543$ & $36390$ & $64042$ & $95245$\\
 						   & Postgres  & $11828$  & $38192$ & $66061$ & $113689$\\\hline\hline
 		\multirow{2}{*}{10K}& PosDB & $12116$ & $38001$ & $67239$ & $100130$\\
 						   & Postgres  & $11909$  & $38460$ & $67449$ & $113798$\\\hline\hline
 		\multirow{2}{*}{100K}& PosDB & $12713$ & $39973$ & $70159$ & $105051$\\
 						   & Postgres  & $11924$ & $38552$ & N/A & N/A\\\hline\hline
 		\multirow{2}{*}{1M}& PosDB & $13272$ & $41552$ & $72982$ & $107273$\\
 						   & Postgres  & N/A & N/A & N/A & N/A \\\hline\hline
 		\multirow{2}{*}{10M}& PosDB & $12693$ & $39580$ & $69677$ & $101774$\\
 						   & Postgres  & N/A & N/A & N/A & N/A \\ \hline
	\end{tabular}
    \end{minipage}%
    \begin{minipage}{0.5\textwidth}
        \centering
	\begin{tikzpicture}[scale=0.75]
	\begin{axis}[
	ymin=0,
	xmin=1,
	legend pos=north west,
	ymajorgrids=true,
	grid style=dashed,
	ylabel shift = -4 em,
	xlabel={Scale Factor},
	ylabel={Time (ms)}
	]	
	\addplot[
	color=Green,
	mark=diamond*
	]
	coordinates {
		(1,10611)(3,32888)(5,57584)(7,84934.7)
	};
	\addplot[
	color=RedOrange,
	mark=pentagon*
	]
	coordinates {
		(1,11498)(3,37205)(5,73003)(7,116160)
	};
	\legend{PosDB with Segment Tree, PostgreSQL}
	\end{axis}
	\end{tikzpicture}
    \end{minipage}
\end{figure}

PosDB and PostgreSQL show approximately equal results on small scale factors (SF): PosDB wins for a small \texttt{@offt}, and PostgreSQL wins for large. Increasing SF leads to increasing advantage of PosDB and increasing \texttt{@offt} allows PostgreSQL to catch up, but not to outperform. At the same time, having a large \texttt{@offt} leads to unresponsive behavior of PostgreSQL (timeout was set to 10 minutes). Increasing SF leads to freezing on smaller window sizes. To the right of the table we present a graph comparing performances of the systems at \texttt{@offt}=10. It demonstrates the benefits of column-stores with late materialization.

Note that since this query favours a sequential scan, we implement it using \textbf{Strategy 1}. A detailed evaluation of \textbf{Strategies 1, 2a}, and \textbf{2b}, as well as assessment of performance impact of ``wide'' join indexes is the subject of future work. Currently, we anticipate that random data accesses spawned by \textbf{Strategies 2a, 2b} threaten to degrade query performance in some cases. However, it depends on a number of parameters: attribute sizes, selectivities of predicates, data distribution, and so on. We believe that at least in a part of cases our approach will still be beneficial, and a proper cost model will highlight it.

\section{Related Work}\label{sec:related_work}

Despite the fact that window functions were proposed almost 20 years ago, there is a surprisingly low number of works on the subject. They can be classified into two groups:

\textbf{Designing the operator itself.} Cao et al.~\cite{Cao:2012:OAW:2350229.2350243} consider a case when a single query contains several window functions. The proposed approach is to reuse grouping and ordering steps. At first, the authors consider two methods of tuple ordering for a single window~--- hashed sort and segmented sort. They discuss their properties and applicability. Finally, they propose an optimization scheme for handling several window functions, which generates an evaluation schedule. Wesley and Xu~\cite{Wesley:2016:ICC:2994509.2994537} propose to reuse the internal state between adjacent frames for computing holistic windowed aggregates. A holistic function is a function that cannot be decomposed using other functions. Therefore, \texttt{MEDIAN} or \texttt{COUNT DISTINCT} are holistic and \texttt{SUM} or \texttt{MIN} are not. Speeding up the evaluation of such window aggregates is a relevant problem since their computation requires looking at all data at once. A paper by Leis et al.~\cite{leis_window_functions_2015} describes an efficient algorithm for the whole window function operator. It considers existing approaches for aggregate computation, as well as proposes a novel one, based on a segment tree. Finally, an efficient parallelization of all steps is discussed.

\textbf{Window functions and external optimization.} Coelho et al.~\cite{10.1007/978-3-319-39577-7-6} addresses reshuffling in a distributed environment for efficient processing of window functions. The authors utilized histograms to assess the size of the prospective groups and their distribution between the nodes. Zuzarte et al.~\cite{Zuzarte:2003:WSE:872757.872840} discuss how and when it is possible to rewrite a correlated subquery using window functions. Such rewriting can significantly improve query performance.

\section{Conclusion}\label{sec:conclusion}

In this study, we have discussed the implementation of window functions in a column-store with late materialization. We have proposed three different strategies, and for each of them we have provided a model for estimating the amount of required memory. We also present an enhancement of the segment tree technique for processing \texttt{RANGE}-based window functions. Experimental comparison with PostgreSQL has demonstrated the viability of this technique.

\bibliographystyle{splncs04}
\bibliography{my}

 \end{document}